\documentstyle[12pt,epsfig]{article}
\begin{document}
\centerline{\Large\bf Schwarzschild Black Hole in Noncommutative Spaces}
\vspace*{0.050truein}
\centerline{Forough Nasseri}
\centerline{\it Department of Physics, Sabzevar University
of Tarbiat Moallem, P.O. Box 397, Sabzevar, Iran}
\centerline{\it Khayyam Planetarium, P.O. Box 769, Neishabour, Iran}
\begin{center}
(\today)
\end{center}

\abstract{We study the effects of noncommutative spaces on the horizon,
the area spectrum and Hawking temperature of a Schwarzschild
black hole. The results show deviations from the usual horizon,
area spectrum and the Hawking temperature. The deviations depend on
the parameter of space/space noncommutativity.}

\section{Introduction}
Recently, remotivated by string theory arguments, noncommutative spaces
(Moyal plane) have been studied extensively. The noncommutative space
can be realized by the coordinate operators satisfying
\begin{equation}
\label{0}
\left[ {\hat x}_{\mu},{\hat x}_{\nu} \right] = i \theta_{\mu \nu},
\end{equation}
where $\hat x$ are the coordinate operators and $\theta_{\mu\nu}$ is the
noncommutativity parameter and is of dimension (length)$^2$;
for a review on the string theory side, see \cite{se}.
In noncommutative spaces, the usual product of fields should be replaced
by the star-product:
\begin{equation}
\label{2}
\left( f \star g \right) =\exp \left( \frac{i}{2}
\theta_{\mu\nu} \frac{\partial}{\partial x^\mu}
\frac{\partial}{\partial y^\nu} \right) f(x) g(y)|_{x=y},
\end{equation}
where $f$ and $g$ are two arbitrary infinitely differentiable functions on
$R^{3+1}$. For noncommutative spacetime ($\theta_{0i} \neq 0$)
it has been shown that the theory is not unitary and hence, as a field
theory, it is not appealing \cite{jg,mc}. 
In noncommutative spaces we have
\begin{equation}
\label{3}
\left[ {\hat x}_i,{\hat x}_j \right]=i \theta_{ij},\;\;\;
\left[ {\hat x}_i,{\hat p}_j \right]=i \delta_{ij},\;\;\;
\left[ {\hat p}_i,{\hat p}_j \right]=0.
\end{equation}
In this letter we focus on the metric of a Schwarzschild black hole
in a noncommutative space. Given the metric and assuming that the
noncommutative parameter ($\theta_{ij}$) is small, we study the
horizon, the area spectrum and the Hawking temperature of the
Schwarzschild black hole.
Since the noncommutativity in space violates rotational symmetry,
our horizon corrections have a preferred direction.
We will use a natural unit system that sets $k_B$,
$c$, and $\hbar$ all equal to $1$, so that $\ell_P=M_P^{-1}=\sqrt{G}$.

The plan of this letter is as follows. In section 2, we present
the horizon, area spectrum and Hawking temperature of
the Schwarzschild black hole in noncommutative spaces. We discuss
our results and conclude in Sec. 3.
\section{Noncommutativity and Schwarzschild black hole}
The metric of the Schwarzschild black hole is given by
\begin{equation}
\label{1}
ds^2= \left( 1 - \frac{2GM}{r} \right) dt^2 -
\frac{dr^2}{\left( 1 - \frac{2GM}{r} \right)}-
r^2 \left( d \theta^2 +\sin^2 \theta d\phi^2 \right).
\end{equation}
There is a horizon at $r_{\rm h}=2GM$. The horizon area $A$ of
the Schwarzschild black hole is given by
\begin{equation}
\label{4}
A=r_{\rm h}^2 \int_{0}^{2\pi} d \phi \int_{0}^{\pi} \sin \theta d \theta=
4 \pi r_{\rm h}^2= 16 \pi G^2 M^2.
\end{equation}
The Hawking temperature of the Schwarzschild black hole is given by
$T_{\rm H}=\kappa/(2 \pi)$ where $\kappa$ is the surface gravity of the
black hole \cite{bi}. For Schwarzschild black hole $\kappa=GM/r_{\rm h}^2$
and therefore
\begin{equation}
\label{5}
T_{\rm H}=\frac{GM}{2 \pi r_{\rm h}^2}=\frac{1}{8 \pi G M}.
\end{equation}
We propose the following metric for the Schwarzschild black hole
in the noncommutative spaces:
\begin{equation}
\label{6}
ds^2= \left( 1 - \frac{2GM}{\sqrt{{\hat r}{\hat r}}} \right) dt^2 -
\frac{d{\hat r}d{\hat r}}{\left( 1 - \frac{2GM}{\sqrt{{\hat r}{\hat r}}} \right)}-
{\hat r}{\hat r} \left( d \theta^2 +\sin^2 \theta d\phi^2 \right),
\end{equation}
where ${\hat r}$ satisfying (\ref{3}).
The horizon of the noncommutative metric (\ref{6}) satisfies
the following condition
\begin{equation}
\label{7}
1 - \frac{2GM}{\sqrt{{\hat r}{\hat r}}}=0.
\end{equation}
Now, we note that there is a new coordinate system,
\begin{equation}
\label{8}
x_i={\hat x}_i+\frac{1}{2} \theta_{ij} {\hat p}_j,\;\;\;p_i={\hat p}_i,
\end{equation}
where the new variables satisfy the usual canonical commutation relations:
\begin{equation}
\label{9}
\left[ x_i,x_j \right] = 0,\;\;\;
\left[ x_i,p_j \right] = i \delta_{ij},\;\;\;
\left[ p_i,p_j \right] = 0.
\end{equation}
So, if in the horizon condition (\ref{7}) we change the variables
${\hat x}_i$ to $x_i$, the horizon of the noncommutative metric (\ref{6})
satisfies the following condition
\begin{equation}
\label{10}
1 - \frac{2GM}{\sqrt{(x_i-\theta_{ij}p_j/2\hbar)(x_i-
\theta_{ik}p_k/2\hbar)}}=0.
\end{equation}
This leads us to
\begin{equation}
\label{11}
1-\frac{2GM}{r}-GM\frac{x_i \theta_{ij} p_j}{r^3}
+{\cal{O}}(\theta^2)=0,
\end{equation}
or
\begin{equation}
\label{12}
1-\frac{2GM}{r}-GM\frac{L . \theta}{2 r^3}
+{\cal{O}}(\theta^2)=0,
\end{equation}
where $\theta_{ij}=\frac{1}{2}\epsilon_{ijk} \theta_k$, $L=r \times p$.
If we put $\theta_3=\theta$ and the rest of the $\theta$-components
to zero (which can be done by a rotation or a redefinition of
coordinates), then $L.\theta=L_z \theta$. So, we can rewrite
Eq. (\ref{12}) as
\begin{equation}
\label{13}
r^3 -2GMr^2 -\frac{GML_z \theta}{2}=0.
\end{equation}
By the following definitions
\begin{equation}
\label{14}
a \equiv -2GM,\;\;\;c \equiv -\frac{GML_z\theta}{2},
\end{equation}
the real root of cubic formula (\ref{13}) is the horizon
of the Schwarzschild black hole in noncommutative spaces
\begin{eqnarray}
\label{15}
{\hat r}_h& \equiv &-\frac{a}{3} 
+ \left(\frac{-2a^3 - 27c + \sqrt{108a^3c+729c^2}}{54}\right)^{1/3}\nonumber\\
&+& \left(\frac{-2a^3 - 27c - \sqrt{108a^3c+729c^2}}{54}\right)^{1/3}.
\end{eqnarray}
Two other roots of cubic formula (\ref{13}) are not real \cite{in}.
In the case of commutative spaces $c=0$, Eq. (\ref{15}) yields
$r_h=2GM$.
The Hawking temperature and the horizon area of Schwarzschild black hole
in noncommutative spaces are respectively
\begin{eqnarray}
\label{16}
T_H &=& \frac{GM}{2\pi {\hat r}_h {\hat r}_h},\\
\label{17}
A &=& 4\pi {\hat r}_h {\hat r}_h.
\end{eqnarray}
Substituting ${\hat r}_h$ from (\ref{15}) into (\ref{16}) and (\ref{17})
give us the Hawking temperature and the horizon area of the
Schwarzschild black hole in noncommutative spaces.
\section{Conclusion}
In this letter, we investigate the effects of noncommutative spaces
on the Schwarzschild black hole. We calculate the horizon, the area
spectrum and the Hawking temperature in the noncommutative spaces.
Our results show deviations from the usual formulae. The deviations
depend on the parameter of space/space noncommutativity.

\end{document}